\documentclass{PoS}

\newcommand{\bda}{\begin{\displaymath}\begin{array}{rl}}
\newcommand{\eda}{\end{array}\end{displaymath}}
\newcommand{\be}{\begin{equation}}
\newcommand{\ee}{\end{equation}}
\newcommand{\bdm}{\begin{displaymath}}
\newcommand{\edm}{\end{displaymath}}
\newcommand{\bea}{\begin{eqnarray}}
\newcommand{\eea}{\end{eqnarray}}

\newcommand{\co}{\; ,}

\newcommand{\ubar}{\overline{\rule[0.42em]{0.4em}{0em}}\hspace{-0.5em}u}
\newcommand{\dbar}{\,\overline{\rule[0.65em]{0.4em}{0em}}\hspace{-0.6em}d}

\newcommand{\ChPT}{$\chi$PT }
\newcommand{\lbar}{\bar{\ell}}

\title{Recent developments in the physics of light quarks }

\ShortTitle{Recent developments in the physics of light quarks }

\author{\speaker{Heinrich Leutwyler}\\
       University of Bern\\
       E-mail: \email{Leutwyler@itp.unibe.ch}}

\abstract{My talk was dedicated to the memory of Jan Stern. The brief account given below focuses on the
progress achieved in the determination of the $\pi\pi$ S-wave scattering lengths, both experimentally
and with light dynamical quarks on a lattice. In view of the excellent agreement, we can conclude
that (a) the expansion in powers of the two lightest quark masses represents a very efficient tool
for the analysis of the low energy structure of QCD and (b) the size of the energy gap of QCD is
governed by the order parameter of lowest dimension, the quark condensate.
\vspace{0.5cm}
\begin{center}
\hspace{-0.4cm}\includegraphics[width=.5\textwidth]{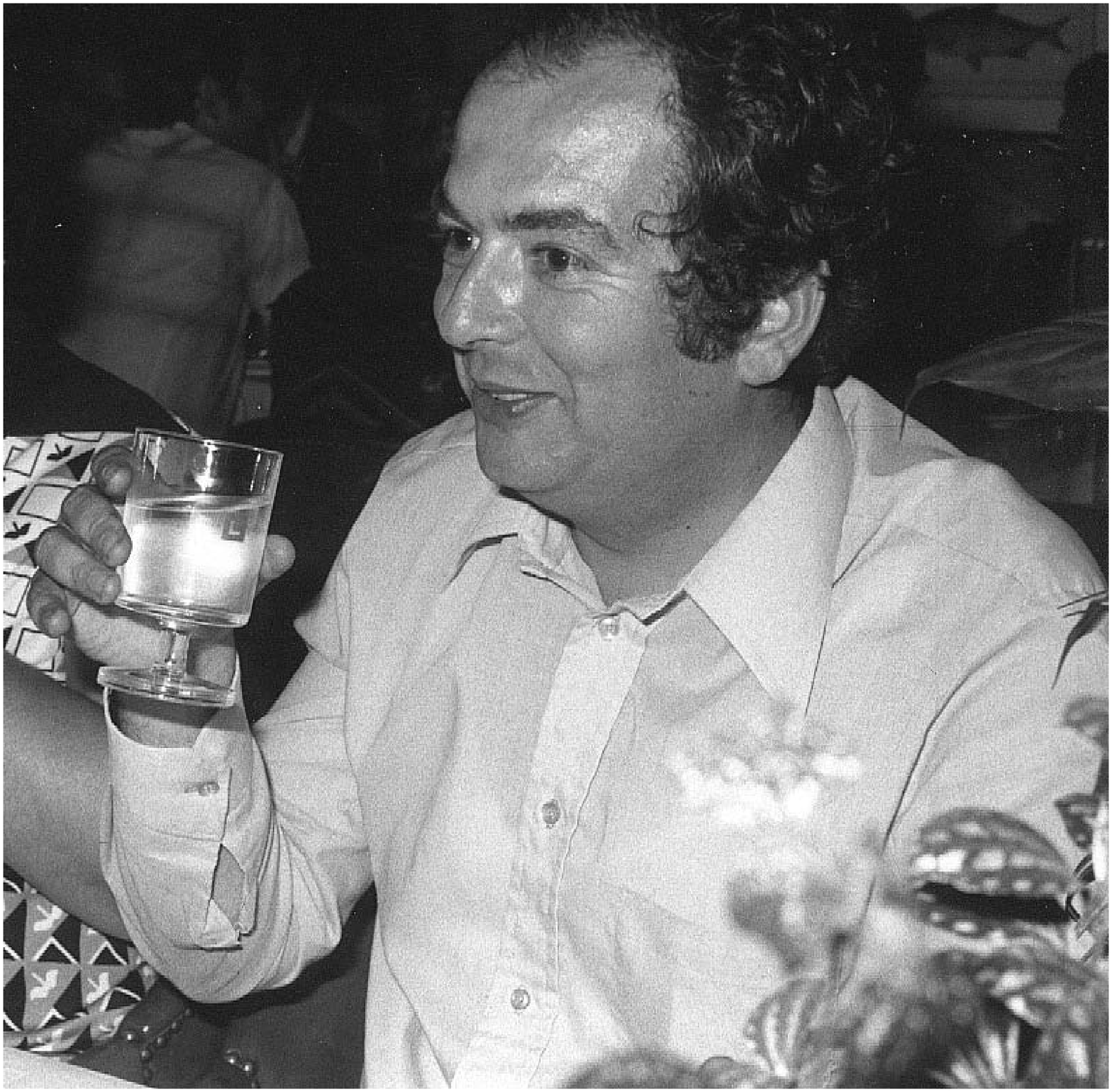}

\vspace{0.5cm}{\it
In memoriam Jan Stern, 29. 6. 1942 -- 2. 7. 2008}\end{center}}

\FullConference{8th Conference Quark Confinement and the Hadron Spectrum\\
		 September 1-6, 2008\\
		 Mainz. Germany}

\begin{document}

\section{Introduction}\label{sec:intro}
In view of the limited space available, the present report only summarizes the first part of my talk, which dealt with the remarkable progress achieved with light dynamical quarks on the lattice and with the low energy precision experiments concerning the $\pi\pi$ S-wave scattering lengths. A more detailed account, which also outlines recent developments in the dispersive analysis of the low energy structure of QCD, is given in \cite{Montpellier 2008}. 

At low energies, the main characteristic of QCD is that the energy gap is remarkably
small, $M_\pi\simeq $ 140 MeV. More than 10 years before the discovery of QCD,
Nambu \cite{Nambu} found out why that is so: the gap is small because the
strong interaction has an approximate chiral symmetry. Indeed, QCD does have
this property: for yet unknown reasons, two of the quarks happen to
be very light. The symmetry is not perfect, but nearly so: $m_u$ and $m_d$ are
tiny. The mass gap is small because the symmetry is ``hidden'' or
``spontaneously broken'': for dynamical reasons, the ground state of the
theory is not invariant under chiral rotations, not even approximately. The
spontaneous breakdown of an exact Lie group symmetry gives rise to strictly
massless particles, ``Goldstone bosons''. In QCD, the pions play this role:
they would be strictly massless if $m_u$ and $m_d$ were zero, because the
symmetry would then be exact. The only term in the Lagrangian of QCD that is not
invariant under the group SU(2)$\times$SU(2) of chiral rotations
is the mass term of the two lightest quarks, $m_u\,\ubar u+m_d \,\dbar
d$. This term equips the pions with a mass. Although the
theoretical understanding of the ground state is still poor, we do have very
strong indirect evidence that Nambu's conjecture is right -- we know why the
energy gap of QCD is small.

\section{Lattice results for $M_\pi$ and $F_\pi$}\label{sec:lat}
As pointed out by Gell-Mann, Oakes and Renner \cite{GMOR}, the square of the
pion mass is proportional to the strength of the symmetry breaking,
$ M_\pi^2\propto (m_u+m_d).$  This property can now be checked on the lattice, where -- in principle -- the quark masses can be varied at will. In
view of the fact that in these calculations, the quarks are treated
dynamically, the quality of the data is impressive. The masses are
sufficiently light for \ChPT to allow a meaningful extrapolation to the
quark mass values of physical interest. The results indicate that the ratio
$M_\pi^2/(m_u+m_d)$ is nearly constant out to values of $m_u, m_d$ that are
about an order of magnitude larger than in nature. According to Gell-Mann, Oakes and Renner, this ratio is related to the quark condensate. The Banks-Casher relation, which connects the quark condensate with the spectral density of the Dirac operator at small eigenvalues \cite{Banks & Casher}, is now also accessible to a numerical evaluation on the lattice \cite{Giusti Mainz}.

The Gell-Mann-Oakes-Renner relation corresponds to the leading term in the expansion of $M_\pi^2$ in powers of $m_u$ and $m_d$  (mass of the strange quark kept fixed at the physical value). At next-to-leading order, the expansion contains a logarithm: 
\be\label{eq:Mpi one loop} M_\pi^2=
M^2\left\{1 +\!\frac{M^2}{32\pi^2 F_\pi^2}\, \ln
  \frac{M^2}{\Lambda_3^2}\!+\!O(M^4)\right\}\co\ee
where $M^2\equiv B(m_u+m_d)$ stands for the term linear in the quark masses. 
Chiral symmetry fixes the coefficient of the logarithm in
terms of the pion decay constant $F_\pi\simeq 92.2$ MeV, but does not determine the scale $\Lambda_3$. An estimate for this scale was obtained more than 20 years
ago \cite{GL SU2}, on the basis of the SU(3) mass formulae for the pseudoscalar
octet: $\lbar_3\equiv \ln \Lambda_3^2/M_\pi^2= 2.9\pm 2.4$. Several collaborations have now managed to determine the scale $\Lambda_3$ on the lattice -- for an overview, I refer to \cite{Ecker & Necco}. The result of the
RBC /UKQCD collaboration, $\lbar_3=3.13 \pm 0.33_{\,\mbox{\footnotesize stat}}\pm 0.24_{\,\mbox{\footnotesize syst}}$, for instance, which concerns 2+1 flavours and includes an estimate of the systematic errors, is perfectly consistent with the number quoted above, but considerably more accurate. 

The expansion of $F_\pi$ in powers of $m_u,m_d$ also contains a logarithm at NLO.  The coupling constant relevant in that case is denoted by $\lbar_4$. A couple of years ago, we obtained a rather accurate result for this quantity, from a dispersive analysis of the scalar form factor: $\lbar_4=4.4\pm0.4$  \cite{CGL} (for details,  I refer to \cite{ACCGL}). The lattice determinations of $\lbar_4$ have reached comparable accuracy and are consistent with the dispersive result \cite{Ecker & Necco}. 

Concerning the expansion in powers of $m_s$, however, the current situation leaves much to be desired. While some of the lattice results indicate, for instance, that the violations of the Okubo-Zweig-Iizuka rule in the quark condensate and in the decay constants are rather modest, others point in the opposite direction. In view of the remarkable progress being made with the numerical simulation of light quarks, I am confident that the dust will settle soon, so that the effective coupling constants that govern the dependence of the various quantities of physical interest on $m_s$  can reliably be determined, to next-to-next-to-leading order of the chiral expansion.

\section{Consequences for the $\pi\pi$ scattering lengths}
 \begin{figure}[thb]\centering
 \includegraphics[width=.5\textwidth,angle=-90]{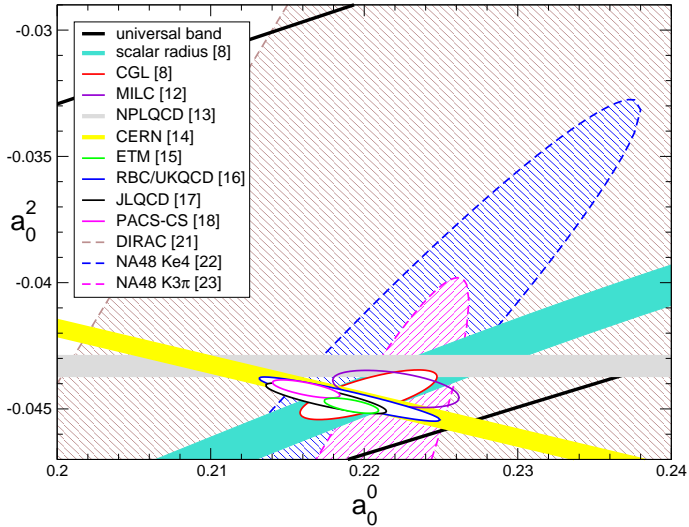}
\caption{\label{fig:a0a2}Lattice results for  $\lbar_3,\lbar_4$  [12-18], converted into S-wave $\pi\pi$ scattering lengths, compared with $K_{e4}$ and $K_{3\pi}$ data and with the prediction in [8], which relied on SU(3) for $\lbar_3$ and on a dispersive evaluation of the scalar radius for $\lbar_4$. Figure prepared in the framework of Flavianet, in collaboration with G.\ Colangelo.}  
\end{figure} 
The hidden symmetry not only controls the size of the energy gap, but also
determines the interaction of the Goldstone bosons at low energies, among
themselves, as well as with other hadrons. In particular, as pointed out by
Weinberg \cite{Weinberg 1966}, the leading term in the chiral expansion of the
S-wave $\pi\pi$ scattering lengths (tree level of the effective theory) is
determined by the pion decay constant. In the meantime, the chiral perturbation series has been worked out to NNLO and, matching the chiral and dispersive representations  of the scattering amplitude, a sharp prediction for the scattering lengths was obtained a couple of years ago: $a_0^0=0.220(5)$, $a_0^2=-0.0444(10)$ \cite{CGL}. The error bars are dominated by the uncertainties in the estimates used for the effective coupling constants $\lbar_3,\lbar_4$, which were quoted above. Since recent work on the lattice has reduced these uncertainties -- particularly in the case of $\lbar_3$ -- the predictions for the scattering lengths have now become even sharper. This is illustrated in Fig.\ \ref{fig:a0a2}, where the various lattice results for  $\lbar_3,\lbar_4$ are converted into results for $a_0^0,a_0^2$, using the formulae in \cite{Leutwyler Chiral Dynamics 2006}. The ellipses indicate the corresponding 1 $\sigma$ contours, obtained by summing up all errors in square, including an estimate for the neglected NNLO corrections in the relation between the scattering lengths and $\lbar_3,\lbar_4$ (some of the lattice results shown concern QCD with $N_f=2$ and are thus subject to an unknown systematic error). Indeed, there are tensions among the lattice results, but the plot shows that all of these are within one standard deviation of the prediction obtained on the basis of \ChPT (red ellipse, taken from \cite{CGL}). Note the scale: the width of the figure corresponds to deviations from the central prediction for $a_0^0$ of less than $10 \%$.  NPLQCD, for instance, quotes the outcome for the exotic scattering length $a_0^2$ to an accuracy of 1\%, systematic errors included  \cite{NPLQCD PR 2008}. The result is obtained by analyzing mixed-action data by means of $\chi$PT to NLO.

\section{Low energy precision experiments}
I add a few remarks concerning the experimental information about the scattering lengths. \\
{\bf 1}. Production experiments such as $\pi N\rightarrow\pi\pi N$, $J/\psi\rightarrow \pi\pi \omega$, $\ldots\,$ provide valuable information about the $\pi\pi$ phase shifts in the intermediate energy region, but since the pions are not produced in vacuo, the analysis is complicated -- the uncertainties in the results for the scattering lengths are much too large for these experiments to be of interest in the present context.\\ 
{\bf 2}. In principle, the decay $K\rightarrow\pi\pi$ can be used to measure the phase difference $\delta^0_0-\delta^2_0$ at the kaon mass. Unfortunately, however, the $\Delta I=\frac{1}{2}$ rule implies that the result for the phase difference is subject to unusually large isospin breaking effects. In the past, work on this problem invariably led to a value for the phase difference that is too large, presumably because isospin breaking was not properly accounted for. Only rather recently, Cirigliano, Ecker, Neufeld and Pich have performed a complete analysis of these transitions, based on \ChPT to NLO  \cite{Cirigliano Kpipi}. Unfortunately, however, the discrepancy persists.  I conclude that, at the present level of our understanding, the uncertainties associated with isospin breaking are too large for the decay $K\rightarrow\pi\pi$ to provide useful information about the low energy structure of QCD.\\
{\bf 3}. The low energy theorem for the scalar radius of the pion correlates the two S-wave scattering lengths to a narrow strip \cite{CGL}. If this correlation is used, together with the corrections for isospin breaking obtained in \cite{CGR}, the $K_{e4}$ data determine $a^0_0$ to the same precision as the theoretical prediction and hit the nail on the head: $a_0^0=0.220(5)(2)$ \cite{Bloch}.   \\
{\bf 4}. As pointed out by Cabibbo \cite{Cabibbo Isidori}, the cusps occurring near threshold in decays of kaons into three pions can be used to measure the combination $a^0_0-a^2_0$ of scattering lengths. A preliminary analysis of the 2003 + 2004 data collected on $K_{3\pi}$ decay at NA48/2 is reported in \cite{Madigozhin}. Using the framework derived in \cite{CGKR}, and the low energy theorem for the scalar radius, these data imply $a_0^0-a^2_0=0.266Ê\pmÊ0.003_{\mbox{\footnotesize stat}}Ê\pmÊ0.002_{\mbox{\footnotesize syst}}Ê\pmÊ0.001_{\mbox{\footnotesize ext}}$, thus subjecting the \ChPT prediction, $a_0^0-a^2_0=0.265\pm0.004$ \cite{CGL}, to a very stringent test.

I conclude that the low energy precision measurements as well as the results obtained on the lattice consolidate the picture developed on the basis of  $\chi$PT: the expansion of the square of the pion mass in powers of $m_u,m_d$ is dominated by the leading term, which is proportional to the quark condensate. The NLO contributions are now known rather accurately -- as expected, they are tiny.


\begin{thebibliography}{99}

\bibitem{Montpellier 2008}H.~Leutwyler,
  %``On the low energy end of the QCD spectrum,''
in Proc.\ {\it QCD 08}, 
Montpellier, France, July 2008, ed.\ S.\ Narison, arXiv:0809.5053. 
  
\bibitem{Nambu}
  Y.~Nambu,
  %``Axial vector current conservation in weak interactions,''
  Phys.\ Rev.\ Lett.\  {\bf 4} (1960) 380.

\bibitem{GMOR}
M.~Gell-Mann, R.~J.~Oakes and B.~Renner,
%``Behavior Of Current Divergences Under SU(3) X SU(3),''
Phys.\ Rev.\ {\bf 175} (1968) 2195.

\bibitem{Banks & Casher}
T.~Banks and A.~Casher,
%``Chiral Symmetry Breaking In Confining Theories,''
Nucl.\ Phys.\  B {\bf 169} (1980) 103.

\bibitem{Giusti Mainz}
L.~Giusti, talk given at this conference.

\bibitem{GL SU2}
J.~Gasser and H.~Leutwyler,
%``Low-Energy Theorems As Precision Tests Of QCD,''
Phys.\ Lett.\ B {\bf 125} (1983) 325;
%``Chiral Perturbation Theory To One Loop,''
Annals Phys.\ {\bf 158} (1984) 142.

\bibitem{Ecker & Necco} G.~Ecker and S.~Necco, talks given at this conference.

\bibitem{CGL} 
  G.~Colangelo, J.~Gasser and H.~Leutwyler,
  %``pi pi scattering,''
  Nucl.\ Phys.\ B {\bf 603} (2001) 125.

\bibitem{ACCGL}
B.~Ananthanarayan {\it et al.},
  %``Scalar form factors of light mesons,''
  Phys.\ Lett.\  B {\bf 602} (2004) 218.

\bibitem{Weinberg 1966}
 S.~Weinberg,
  %``Pion Scattering Lengths,'' 
  Phys.\ Rev.\ Lett.\  {\bf 17} (1966) 616.
  
\bibitem{Leutwyler Chiral Dynamics 2006}H.~Leutwyler, %$\pi \pi$ scattering, 
in Proc.\ {\it Chiral Dynamics, Theory \& Experiment}, Durham/Chapel Hill, NC, USA,
eds.~M.~W.~Ahmed et al., World Scientific, Singapore (2007), p.17, hep-ph/0612112.

\bibitem{MILC}C.~Bernard {\it et al.} [MILC Collaboration],    
%``Status of the MILC light pseudoscalar meson project,''
 PoS {\bf LAT2007} 090.

\bibitem{NPLQCD PR 2008} S.~R.~Beane et al. [NPLQCD Collaboration], 
  %``Precise Determination of the I=2 pipi Scattering Length from Mixed-Action
  %Lattice QCD,''
  Phys.\ Rev.\  D {\bf 77} (2008) 014505;\\ 
  A.~Walker-Loud, talk given at this conference.

\bibitem{CERN}
  L.~Del Debbio, L.~Giusti, M.~L\"uscher, R.~Petronzio and N.~Tantalo,
  %``QCD with light Wilson quarks on fine lattices. I: First experiences and
  %physics results,''
  JHEP {\bf 0702} (2007) 056.

\bibitem{ETM}
  Ph.~Boucaud {\it et al.}  [ETM Collaboration],
  %``Dynamical twisted mass fermions with light quarks,''
  Phys.\ Lett.\  B {\bf 650} (2007) 304. 

\bibitem{RBC/UKQCD}
  C.~Allton {\it et al.}  [RBC-UKQCD Collaboration],
  %``Physical Results from 2+1 Flavor Domain Wall QCD and SU(2) Chiral
  %Perturbation Theory,''
  arXiv:0804.0473.

\bibitem{JLQCD}
  J.~Noaki {\it et al.}  [JLQCD and TWQCD Collaborations],
  %``Convergence of the chiral expansion in two-flavor lattice QCD,''
  Phys.\ Rev.\ Lett.\  {\bf 101} (2008) 202004.

\bibitem{PACS}
 D.~Kadoh {\it et al.}  [PACS-CS Collaboration],
  %``Application of chiral perturbation theory to 2+1 flavor lattice QCD with
  %O(a)-improved Wilson quarks,''
  PoS  {\bf LAT2007} 109.

\bibitem{Cirigliano Kpipi}V.~Cirigliano, G.~Ecker, H.~Neufeld and A.~Pich,
  %``Isospin violation in epsilon',''
  Phys.\ Rev.\ Lett.\  {\bf 91} (2003) 162001;
  %``Isospin breaking in K --> pi pi decays,''
  Eur.\ Phys.\ J.\  C {\bf 33} (2004) 369.
  
\bibitem{Flavianet Kpipi}
 V.~Cirigliano, C.~Gatti, M.~Moulson, M.~Palutan, for the FlaviaNet Kaon Working Group, 
 % ``pi pi Phase shifts from K to 2 pi, ''
arXiv:0807.5128.

\bibitem{DIRAC}
 B.~Adeva {\it et al.}  [DIRAC Collaboration],
  %``First measurement of the pi+ pi- atom lifetime,''
  Phys.\ Lett.\ B {\bf 619} (2005) 50.
   
\bibitem{CGR} G.~Colangelo, J.~Gasser and A.~Rusetsky,
  %``Isospin breaking in Kl4 decays,''
  arXiv:0811.0775;

\bibitem{Bloch}B.~Bloch-Devaux, talk given at this conference.

\bibitem{Cabibbo Isidori}N.~Cabibbo,
  %``Determination of the a0-a2 pion scattering length from K->pi+ pi0 pi0 decay,''
  Phys.\ Rev.\ Lett.\  {\bf 93} (2004) 121801; N.~Cabibbo and G.~Isidori,
  %``Pion-pion scattering and the K->3pi decay amplitudes,''
  JHEP {\bf 0503} (2005) 021.

\bibitem{Madigozhin}D.\ Madigozhin, talk given at Flavianet Kaon Workshop, Anacapri 2008.

\bibitem{CGKR}G.~Colangelo, J.~Gasser, B.~Kubis and A.~Rusetsky,
  %``Cusps in K --> 3pi decays,''
  Phys.\ Lett.\  B {\bf 638} (2006) 187.

\end{thebibliography}
\end{document}